\documentclass{elsart}
\usepackage{epsfig}
\usepackage{amsmath}

\usepackage{amssymb}
\newcommand{\sn}{\mathop{\mbox{sn}}}
\newcommand{\dn}{\mathop{\mbox{dn}}}
\newcommand{\cn}{\mathop{\mbox{cn}}}
\newcommand{\cE}{\mathcal{E}}

\begin{document}

\def\myint{\int \!\! d^{ {\scriptscriptstyle D} } \boldsymbol{x} \; }
\begin{frontmatter}

\title{Localized and periodic exact solutions to the nonlinear Schr\"odinger equation with spatially modulated parameters: Linear and nonlinear lattices.}

\author{Juan Belmonte-Beitia$^{a}$, Vladimir V. Konotop$^{b}$, V\'{\i}ctor M. P\'erez-Garc\'{\i}a$^{a}$}
\author{Vadym E. Vekslerchik$^{a}$}
\address{$^a$Departamento de Matem\'aticas, E. T. S. de Ingenieros Industriales and Instituto de Matem\'atica Aplicada a la Ciencia y la Ingenier\'{\i}a (IMACI), \\
Universidad de Castilla-La Mancha 13071 Ciudad Real, Spain.}
\address{$^b$Departamento de F\'{\i}sica, Faculdade de Ci\^encias,
Universidade de Lisboa, Campo Grande, Ed. C8, Piso 6, Lisboa
1749-016, Portugal and Centro de F\'{\i}sica Te\'{o}rica e Computacional, Universidade de Lisboa,
 Complexo Interdisciplinar, Avenida Professor Gama Pinto 2, Lisboa
1649-003, Portugal.}




\begin{abstract}
Using similarity transformations we construct explicit solutions of the nonlinear Schr\"odinger equation with linear and nonlinear periodic potentials. We present explicit forms of spatially localized and periodic solutions, and study their properties. We put our results in the framework of the exploited perturbation techniques and discuss their implications on the properties of associated linear periodic potentials and on the possibilities of stabilization of  gap solitons using polychromatic lattices.
\end{abstract}

\end{frontmatter}

\section{Introduction}

The significant increase of the interesting properties of the solutions of the nonlinear Schr\"odinger (NLS) equation with a periodic potential during the last decade was strongly stimulated by the application of that model in the theory of Bose-Einstein condensates (BECs) (see e.g.~\cite{BK_rev,reviews}), where the model is known also as the Gross-Pitaevskii equation. The major advantage of the use of optical lattices, which constitute the physical origin of the periodic potential, is the possibility of change of their characteristics in space what stimulated studies of managing soliton dynamics by spatially inhomogeneous lattices~\cite{BKK}, by low-dimensional lattices~\cite{BMS}, by lattices with defects~\cite{BKPG,BKPG1,BKPG2}, etc. 

Independently recently there has been interest on the systematic study of nonlinear lattices~\cite{SM,Fibich,Sivan2} in a more general context, including not only the BEC mean-field theory but also electromagnetic wave propagation in layered Kerr media. Subsequent studies of the interplay between linear and nonlinear lattices~\cite{BludKon,interplay} have revealed more peculiarities of localized modes in such structures and unusual stability properties of plane wave solutions.

Search of the localized modes in all the studies mentioned above was carried out within the framework of approximate and numerical methods, as the  models, in general case, do now allow for exact solutions. However, recently it was realized that for a number of NLS models including specific type of  periodic potentials, exact periodic solutions can be constructed~\cite{VarPer,VarPer2}, 
what can be done in a systematic manner, using a kind of inverse engineering and starting with a given periodic field distribution~\cite{BK_rev}. We want also to mention the Refs. \cite{JVV,CSF1,CSF2,CSF3}, where different methods to find exact solutions to nonlinear Schr\"odinger equations were studied.

Even greater progress was achieved in constructing exact solutions, and in particular of integrable models, of the NLS equation with time dependent coefficients (see e.g.~\cite{Serkin,Serkin1,Serkin2,Serkin3,Serkin4,Serkin5,Serkin6} for the work devoted to equations of the NLS-type), and for the stationary GP equation with a potential and varying nonlinearity~\cite{exact,exact2}. However, all mentioned models allowing for exact solutions look sometimes rather artificial, for example models with exact periodic solutions must be infinite or subject to cyclic boundary conditions and models with trap potentials require specific laws of time variations of the coefficients. Despite those difficulties, they have two important advantages the first being that they are still experimentally feasible,  and the second being that they allow for exact solutions, thus  ruling out any ambiguity in interpretation of the physical phenomena. 

Returning to the NLS models with spatially periodic coefficients, we observe that no exact  {\em localized} solutions have been reported, so far (the only work to our knowledge is the analytical approach for constructing sufficiently narrow pulses, recently elaborated in~\cite{Sivan}). This is the main goal of this present paper to present for the first time such solutions, which can be obtained for specific types of linear and nonlinear lattices. Moreover, simple analysis of the presented models and their solutions will allow us to make several conclusions about properties of the underlying linear lattices and about stabilization of localized modes by polychromatic lattices. The results presented in this work, represent the extension of the ideas elaborated in earlier publications~\cite{exact,exact2} to the case of periodic nonlinearities. 
 
The paper is organized as follows. In Section 2, we develop the theory of similarity transformations for our model problem: the nonlinear Schr\"odinger equation with a spatially inhomogeneous nonlinearity. In Section 3, we study the stationary localized modes of the inhomogeneous nonlinear Schr\"odinger equation (INLSE). In Section 4, we present a study of the stability of such solutions. Finally, in Section 5, we construct explicit periodic solutions of the INLSE and study their stability.

\section{Similarity transformations}

We consider the one-dimensional spatially inhomogeneous NLS equation  
\begin{equation}
  i \psi_{t} = - \psi_{xx} + v(x) \psi + g(x) \left| \psi \right|^{2} \psi,  
\label{eq-physical1}
\end{equation}
with $x\in\mathbb{R}$, $v(x)$ and $g(x)$ being respectively linear and nonlinear periodic potentials, whose periods will be required to be equal. More specifically, without loss of generality we impose the period to be $\pi$: i.e. $v(x+\pi)=v(x)$ and $g(x+\pi)=g(x)$, what can be always achieved by proper rescaling of variables. (It is worth to mention here that in the BEC theory the choice of the period $\pi$ corresponds to the scaling where the energy is measured in the units of the recoil energy). In order to eliminate possible unessential constant energy shifts and bring the statement of the problem closer to the standard one we impose the requirement for the periodic potential to have a zero mean value, i.e. 
\begin{equation}
\label{mean}
\langle v(x)\rangle \equiv \frac 1\pi\int_{0}^\pi v(x)dx=0.
\end{equation}
The complex field $\psi(t,x)$ will be referred to as a (macroscopic) wave function, where again we bear in mind BEC applications.

In this paper, we focus on stationary solutions of Eq. (\ref{eq-physical1}), which are of the form 
\begin{equation}
\psi(t,x)=\phi(x) e^{-i\mu t},
\end{equation}
where $\phi(x)$ is a function of $x$ only and $\mu$ is a constant below referred to as a chemical potential. As it is clear 
\begin{equation}
\label{eq-physical}
-\phi_{xx}+(v(x)-\mu)\phi+g(x)\phi^{3}=0.
\end{equation}

Now, following the ideas of~\cite{exact,exact2}, we look for a transformation reducing Eq. (\ref{eq-physical}) to the solvable stationary NLS equation
\begin{equation}
  E \Phi = - \Phi'' + G \left| \Phi \right|^{2} \Phi,
\label{eq-stationary}
\end{equation}
where $\Phi \equiv \Phi(X)$, a prime stands for the derivative with respect to $X$, $E$ and $G$ are constants, and $X\equiv X(x)$ is a new spatial variable. To this end, we use the ansatz
\begin{equation}
\label{psi-from-Phi}
\phi(x)=\rho(x)\Phi(X),
\end{equation}
where $\Phi(X)$ is a solution of the stationary equation (\ref{eq-stationary}) and both $\rho(x)$ and $X(x)$ are functions which must be found from the condition that $\psi(x)$ solves Eq. (\ref{eq-physical}). 

Substituting (\ref{psi-from-Phi}) into (\ref{eq-physical}) we obtain the link:
\begin{eqnarray}
    \left( \rho^{2} X_{x} \right)_{x}=0,       
\label{syst}
\end{eqnarray}
as well as expressions for $g(x)$  and $v(x)$ through $\rho(x)$ and $X(x)$:
\begin{eqnarray}
\label{g-one}
 g(x) =  G \frac{X^{2}_{x}}{\rho^{2}}, \quad \text{and}\quad
  v(x) =   \frac{ \rho_{xx} }{ \rho }+\mu 
    - E X_{x}^{2}.
\label{v-one}
\end{eqnarray}
 From the expression for $g(x)$ we find the first constraint of the theory: the method is applicable to models with sign definite nonlinearities, and one must choose $G=$sign$\,(g(x))$. Next, being interested in solutions existing on the whole real axis and excluding singular potentials we have to restrict the analysis to functions $\rho(x)$ in $C^2(\mathbb{R})$ which do not acquire zero values and thus are sign definite. Moreover since neither (\ref{g-one}) nor (\ref{syst}) change as $\rho(x)$ changes the sign (what simply reflects the phase invariance of the model (\ref{eq-physical})), in what follows we restrict the consideration to the case $\rho(x)>0$.

The solution of (\ref{syst}) is immediate:  
\begin{eqnarray}  
\label{rho}
 X(x)=\int_{0}^{x}\frac{ds}{\rho^{2}(s)}.
\label{x_fin}
\end{eqnarray}
Here we have taken into account that the constant of the first integration with respect to $x$ can be chosen 1, as far as $\rho(x)$ is still left undefined, and the second integration constant  conveniently fixes the origin. Thus we have 
\begin{eqnarray}
\label{limits}
X(0)=0,\quad\mbox{and}\quad \lim_{x\to\pm\infty}X(x)=\pm\infty\,.
\end{eqnarray}

Now the coefficients in Eq. (\ref{eq-physical}) can be rewritten in the form
\begin{eqnarray}
\label{rho_fin}
\label{vg_fin}
  v(x)  =  
  \frac{ \rho_{xx} }{ \rho } 
    -  \frac{ E }{ \rho^{4} }+\mu,
 \qquad  g(x)  =  \frac{G }{\rho^{6}}.
\end{eqnarray}

Thus, (\ref{psi-from-Phi}) with (\ref{rho}) gives a solution of Eq. (\ref{eq-physical}), which depends on the  the positive definite function $\rho(x)$, which must be chosen $\pi$-periodic. This choice is naturally very rich. To be specific we investigate in detail the simplest case where
\begin{equation}
\label{eleccion1}
\rho(t,x)=1+\alpha \cos(2x), 
\end{equation}
with $\alpha$ being a real constant, such that $|\alpha|<1$, and respectively 
\begin{eqnarray}
\label{v_per} 
v(x)&=&
-\frac{4\alpha \cos(2x)}{1+\alpha\cos(2x)}-\frac{E}{(1+\alpha\cos(2x))^{4}} +\mu,
\\
\label{g_per}
g(x)&=&\frac{G}{(1+\alpha\cos(2x))^{6}},
\end{eqnarray}
where $\mu$ is chosen to ensure (\ref{mean}) and reads 
\begin{equation}\label{cp}
\mu= 
4-\frac{4}{\sqrt{1-\alpha^2}}+E\frac{2+3\alpha^{2}}{2(1-\alpha^{2})^{7/2}}.
\end{equation}
Some direct generalizations of this model are presented in the Appendix.

The introduced linear periodic potential $v(x)$ has a number of peculiar properties, some of which are:
 
\begin{quotation}
 
	(i) It allows for some particular explicit forms of the Bloch function (see e.g. expressions (\ref{Bloch_10}) and (\ref{Bloch_exact}), below), what is a rather peculiar situation since in a general situation explicit forms of solutions of a Hill equation in terms of the elementary fucntions are not available.

\smallskip
	
\noindent 	
	(ii) $\alpha$ appears to be the parameter allowing one to control the band spectrum (as this is illustrated in each of the panels in Fig.~\ref{Math} below) and, in particular, allowing for smooth transition between continuum spectrum (at $\alpha=0$ where the periodic potential does not exist   and the discrete spectrum (at $|\alpha|=1$ where the periodic potential is represented by a sequence of potential wells of infinite depths). In  Fig.~\ref{Math} these two situations are expressed by the fact that all gaps are collapsed at $\alpha=0$, one the one hand, and one of the bands acquire an infinite width at $\alpha=1$, on the other hand.

\smallskip

\noindent
	(iii) By varying the parameter $E$ one can transform the potential from a general form with no intervals of {\em coexistence} ($E<0$) to a form with {\em coexistence} ($E>0$) . Hereafter,  following~\cite{MW} under the coexistence phenomenon we understand the occurrence when an interval of instability (i.e. a gap) disappears due to collapsing of two gap edges. This property is also illustrated in  Fig.~\ref{Math}, where panels (a) and (b) correspond to the general situation while in the panes (c) and (d) one observes intersection of a dashed-doted (red) line with the gap edges precisely in the co-existence points.
 
\end{quotation} 
 
It is remarkable that these properties become more clear departing from the solution of the respective nonlinear problem.  We consider them in the next sections.

\section{Stationary localized modes}

We start with the analysis of the localized modes. To this end we impose the boundary conditions
$\lim_{x\to\pm\infty}\phi(x)=0$ and taking into account (\ref{psi-from-Phi}) and (\ref{limits}) we conclude that $\Phi(X)$ must satisfy the zero boundary conditions, as well: $\lim_{X\to\pm\infty}\Phi(X)=0$. This is the case where $G=-1$ and $\Phi(X)=\sqrt{-2E}/\cosh(\sqrt{-E} X)$ 
i.e. $\Phi(X)$ is the standard stationary NLS soliton. The  respective solution of Eq. (\ref{eq-physical}) reads
\begin{equation}
\label{eq-sol}
\phi_s(x)=\sqrt{-2E}\frac{1+\alpha\cos(2x)}{\cosh(\sqrt{-E} X(x))},\qquad X(x)=\int_0^x\frac{ds}{(1+\alpha\cos(2s))^2}.
\end{equation}
  
Let us now consider in more detail the obtained solution. First of all we recall that the chemical potential of any stationary solution of the NLS equation with a periodic potential, which tends to zero or acquires a  zero value must belong to a forbidden gap of the respective potential and such solutions have a space-independent phase, and thus can be chosen real (see e.g.~\cite{BK_rev,AKS}). The obtained solution (\ref{eq-sol}) has chemical potential $\mu$, given by Eq. (\ref{cp}), and hence it must belong to a gap of the spectrum of the potential $v(x)$, which is determined by the Hill eigenvalue problem: 
\begin{equation}
\label{Hill}
- \varphi_{xx}+v(x)\varphi=\mathcal{E} \varphi. 
\end{equation}  
Moreover, since the obtained solution (\ref{eq-sol}) exists for arbitrary negative $E$, including the limit where $E\to -\infty$ and consequently $\mu< v(x)$ one concludes that $\cE=\mu$ belongs to the semi-infinite gap $(-\infty, \cE_1^{(-)})$ of the band spectrum of $v(x)$, i.e. $\cE_1^{(-)}>\mu$, where we use the notations $\cE_n^{(-)}$ and $\cE_n^{(+)}$ for the lower and upper edges of the $n$-th band, i.e. stability region, provided that $\alpha=1$ designates the lowest band. Indeed, assuming the opposite, i.e. assuming that at some negative energy the solution belongs to one of the finite gap and taking into account the continuous dependence on $E$ one concludes that at some negative $E$, the chemical potential (\ref{cp}) crosses the stability region, what contradicts to the existence of the solution with zero boundary conditions. Alternatively, the above conclusion follows form the facts that at $\alpha=0$ the solution (\ref{eq-sol}) is in the semi-infinite gap and that the dependence of $\mu$ and $v(x)$ on the parameter $\alpha\in (-1,1)$ is continuous. 

\begin{figure}
\epsfig{file=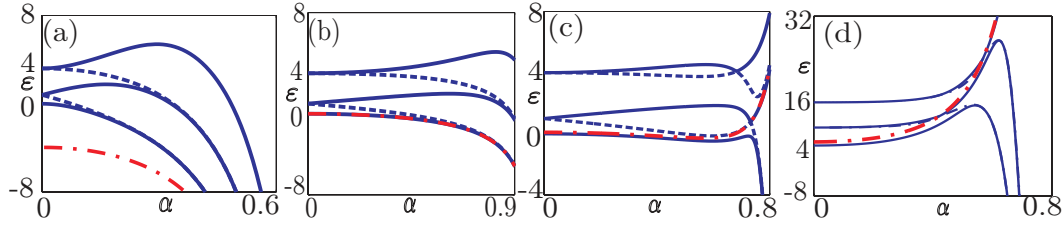,width=14cm}
\caption{Boundaries of the lowest band edges (5 first band edges) of the spectrum of Eq. (\ref{Hill}) as functions of $\alpha$ for
(a) $E=-5$ (b) $E=0$, (c) $E=0.1$, and (d) $E=1$. Solid and dashed lines correspond to $\cE_n^{(-)}$ and to $\cE_n^{(+)}$. The (red) dashed-dot lines represent the chemical potential $\mu$. }
\label{Math}
\end{figure}

The described phenomenon is illustrated in Fig.~\ref{Math} (specifically in the panel (a)) where we show the edges of the lowest bands {\it vs} the parameter $\alpha$ for four typical situations $E<0$ (panel (a)); $E=0$ (panel (b)); $E\in (0,E_0)$ with $E_0=\max_{\alpha}\{E_m(\alpha)\}=2/5$ and $E_m(\alpha)=8(1-\alpha^2)^3/(20+15\alpha^2)$ being the point of the local minimum of the chemical potential $\left. \partial\mu/\partial E\right|_{E=E_0}=0$ for a given $\alpha$, (panel (c)), and $E>E_0$ (panel (d)).  

Leaving the discussion of the situations depicted in Fig.~\ref{Math} (c), (d) for Sec.~\ref{sec:periodic}, now we turn to the limiting cases of the solution (\ref{eq-sol}). At $E\to-\infty$ it approaches the NLS soliton: $\phi_s(x)\sim\sqrt{-2E}(1+\alpha)/\cosh(\sqrt{-E} (1+\alpha)x)$, what reflects the fact that the localization region, determined by $1/\sqrt{|E|}$, is much smaller than the period of the potential. Thus (\ref{eq-sol}) is an example of an exact solution, whose approximate form can be obtained as suggested in \cite{Sivan} (after the scaling out the amplitude $\sqrt{-E}$).

\begin{figure}
\epsfig{file=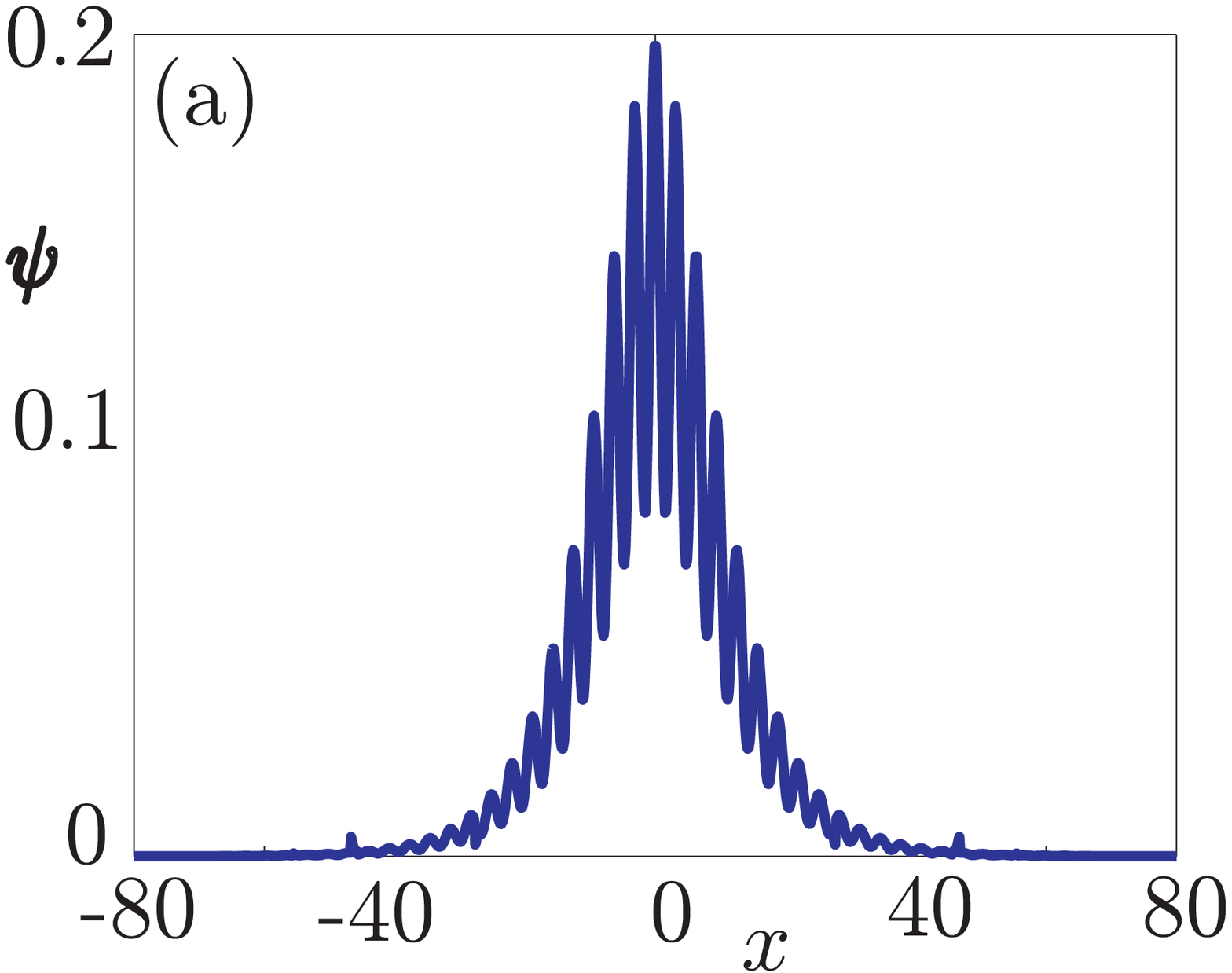, width=7cm}
\epsfig{file=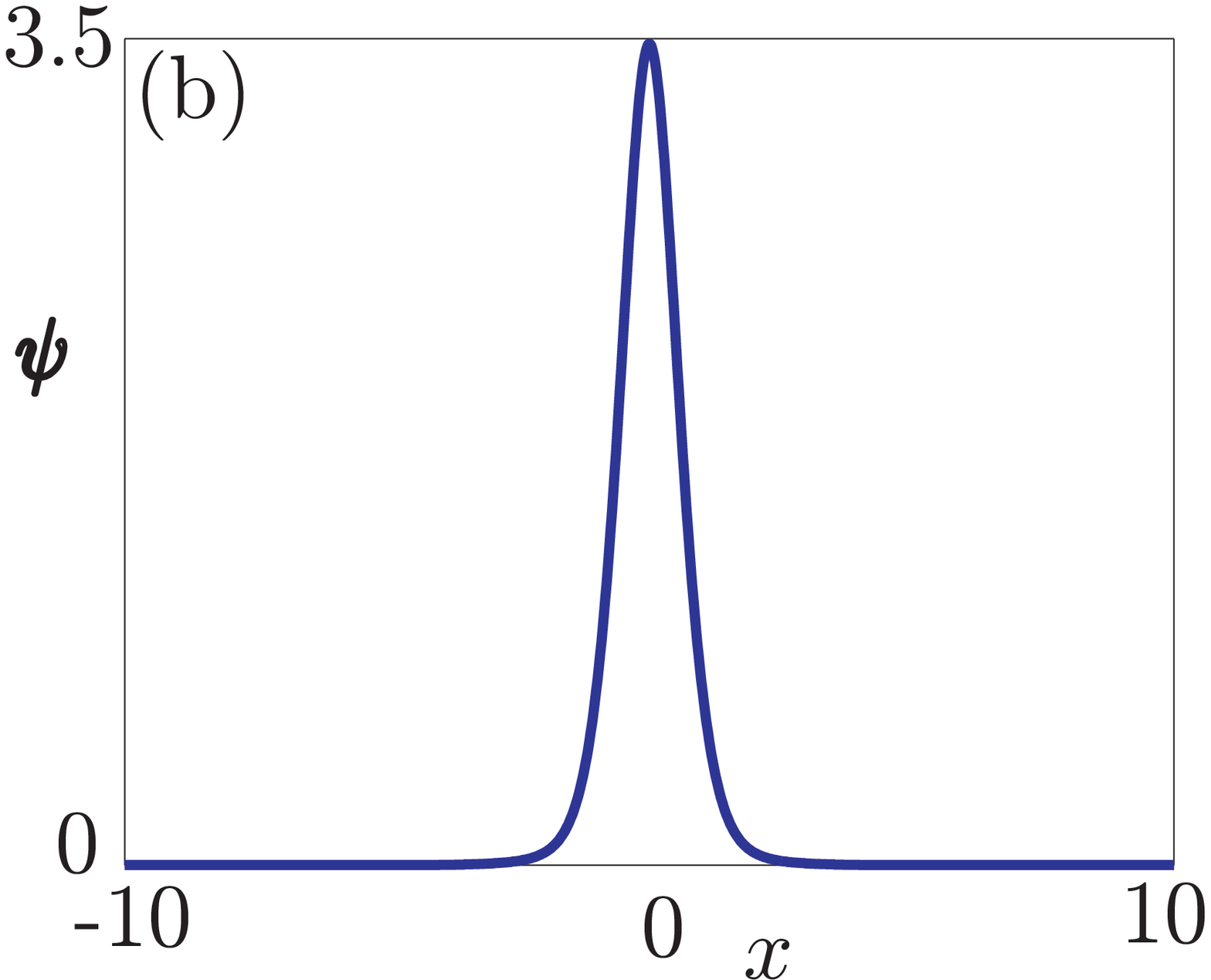,width=7cm}
\caption{
Solutions of Eq. (\ref{eq-physical}), which are given by (\ref{eq-sol}), for (a) $E=-0.01$, $\alpha=0.3$ and (b) $E=-5$, $\alpha=0.1$.
\label{limitcase}}  
 \end{figure}

An even more interesting situation corresponds to small $|E|$. When $|E|\to 0$ the chemical potential approaches the band edge, what is illustrated in Fig.~\ref{Math} (b), where $\cE=\mu$ coincides with the edge of the semi-infinite band (in other words, here we are dealing with a situation where a formula for dependence of the lowest band edge on the parameters of the problem is given explicitly by (\ref{cp})). As it is well known (see e.g. \cite{BK_rev,AKS} and references therein) in this case the solution is accurately described by the multiple-scale approximation and represents an envelope of the Bloch state corresponding to the gap edge. The structure of (\ref{eq-sol}) implies that $\Phi(X)$ is the envelope while 
\begin{eqnarray}
\label{Bloch_10}
\varphi (x)=1+\alpha\cos (2x),
\end{eqnarray}
is the {\em exact} Bloch function  of the potential $v(x,0)$, given by (\ref{v_per}), which corresponds to the lowest edge of the first band. 
%
In other words, in the limit $|E|\to 0$, (\ref{eq-sol}) is an {\em exact envelope soliton solution} bifurcating from the edge of the linear spectrum (to the best of authors knowledge (\ref{eq-sol}) is the  first known solution of such a type). In Fig. \ref{limitcase} we illustrate the two opposite limits of (\ref{eq-sol}) corresponding to an envelope (panel a) and to a narrow (panel b) NLS-type soliton.

Finally we observe that by changing the parameter $\alpha$, one can scan the semi-infinite gap, obtaining the localized mode with {\it a priori} given detuning towards the gap.
 
\section{Stability of the solutions}
 

To check the linear stability of the solution (\ref{eq-sol}) we study the evolution of small perturbations of the form $\psi(x,t) = \left( \phi(x) + f(x,t)+ih(x,t)\right)$, where $f$ and $h$ are real functions,
which leads to the standard linearized Schr\"odinger equation
\begin{equation}
\partial_{t}\left(
\begin{array}{c}
f \\
h
\end{array}
\right)=N\left(
\begin{array}{c}
f \\
h
\end{array}
\right),
\end{equation}
%
with
\begin{equation}
N=\left(
\begin{array}{cc}
0 & L_{-}\\
-L_{+} & 0
\end{array}
\right),
\end{equation}
and
\begin{eqnarray}
\label{operadorLmenos}
L_-=  - \partial_{xx}+v(x)   + g(x) \phi^2(x),
\\ \label{operadorLmas}
L_+ = - \partial_{xx}+v(x)  + 3g(x) \phi^2(x),
\end{eqnarray}

For perturbations $f, h\propto e^{i\Omega t}$, we have
\begin{equation}
\label{prob-autov}
\Omega^{2}f=L_{-}L_{+}f.
\end{equation}
The operators $L_{-}$ and $L_{+}$ are self-adjoint. In the following, we will study some properties of these operators. Using the fact that $\phi$ satisfies the Eq. (\ref{eq-physical}), one easily checks that the operator $L_{-}$ can be rewritten as
\begin{equation}
L_{-}=-\frac{1}{\phi}\partial_{x}\left(\phi^2\partial_{x}\left(\frac{1}{\phi} \cdot\right)\right).
\end{equation}
As a consequence, $\int fL_{-}fdx=\int |\partial_{x}\left(\frac{f}{\phi}\right)|^{2}\phi^{2}dx\ge 0$, and the operator $L_{-}$ is nonnegative. Thus, the spectrum of the operator $L_{-}$ is composed of:
\begin{quotation}

(i) A simple eigenvalue $\lambda_{-}=0$, with the corresponding even eigenfunction, which is solution of Eq. (\ref{eq-physical}).

 \smallskip
 
 \noindent
(ii) A strictly positive continuous spectrum $[\beta,\infty)$.

\end{quotation}
With regard to $L_{+}$ operator, it satisfies the following relation $L_{+}=L_{-}+2g(x) \phi^2(x)$. As $g(x)$ is a negative function, the first eigenvalue of $L_{+}$, $\lambda_{+}$, satisfies $\lambda_{+}<0$. So, at least, an eigenvalue of $L_{+}$ is negative.

Thus, as the $L_{-}$ operator is nonnegative but the $L_{+}$ operator is nonpositive, the composition $L_{-}L_{+}$ is indefinite. Thus, we can not say anything analytically of the sign of $\Omega^{2}$. Moreover, the fact that the potential (\ref{v_per}) can take both positive and negative values complicates further the analytical study.  Therefore, we have to resort to numerical methods to solve Eq. (\ref{prob-autov}). 

\begin{figure}
\epsfig{file=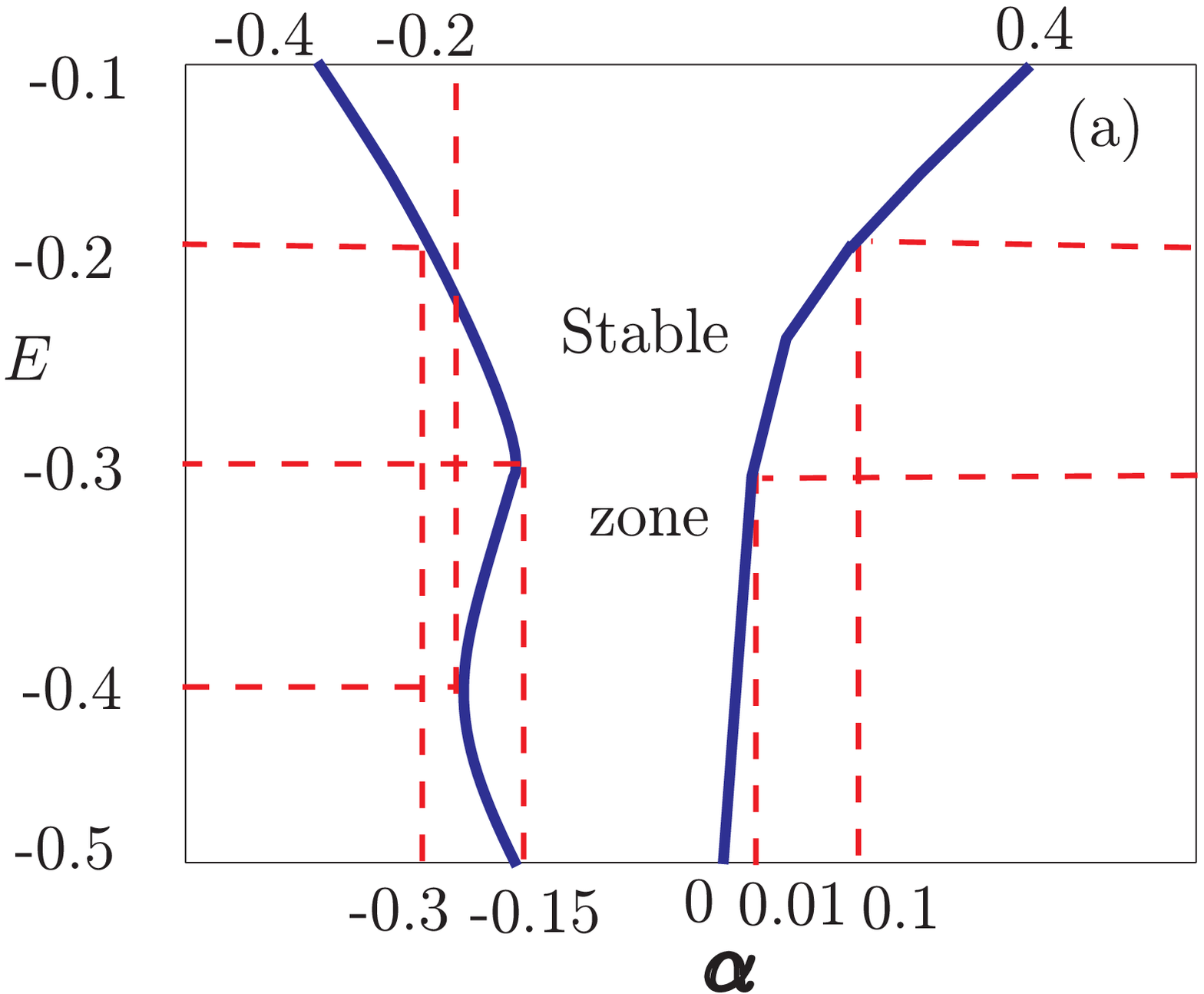,width=7cm}
\epsfig{file=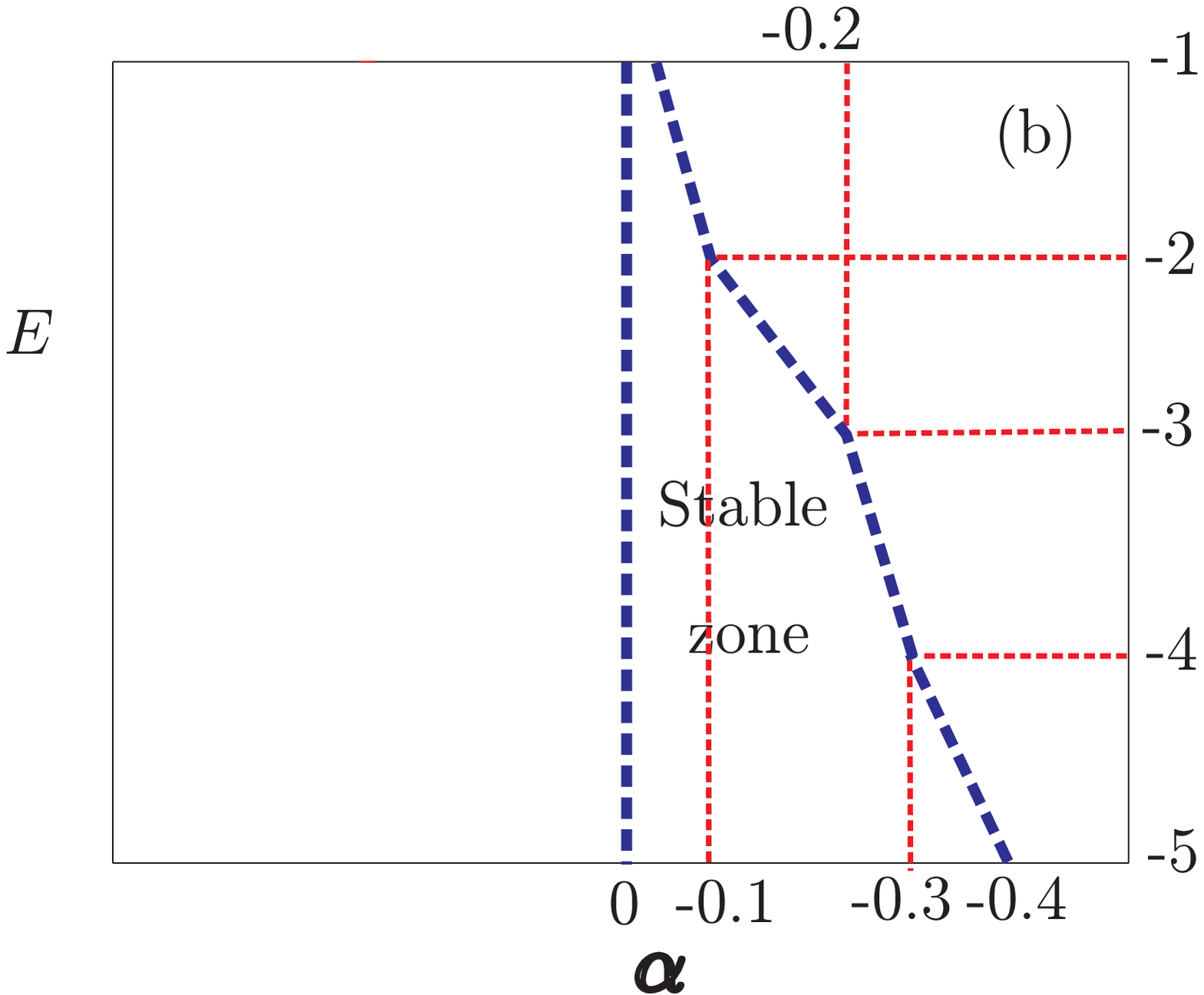,width=7cm}
\caption{[Color online]  Stable zones of the solutions (\ref{eq-sol}), for different values of the parameters $\alpha$ and $E$ (a) $-0.5\leq E\leq-0.1$ (b) $-5\leq E\leq -1$. 
\label{autovalores}}
\end{figure}

 If some eigenvalue $\Omega^{2}$ is negative, the associated solution is unstable. Otherwise, the solution is stable. We can calculate the eigenvalues of the operator $L_{-}L_{+}$ numerically through a direct discretization of the $L_{-}L_{+}$ operator. In Figs. \ref{autovalores} (a) and (b), we show the stable  zones of the solutions (\ref{eq-sol}), for different values of the parameters $\alpha$ and $E$. In Fig. \ref{autovalores} (a), we have studied the case $-0.5\leq E\leq-0.1$. The zones of stability for the case $-5\leq E\leq -1$ are shown in Fig. \ref{autovalores}(b). 
To confirm this result, we have studied numerically the evolution of the solutions (\ref{eq-sol}) under finite amplitude perturbations. We have confirmed that these obtained solutions are stable in the range given by Figs. \ref{autovalores} (a) and (b). 

\section{Exact quasi-periodic solutions}
\label{sec:periodic}

Let us now turn to (quasi-) periodic solutions of Eq. (\ref{eq-physical}). To this end, first, we have to recall the solutions of the NLS equation (\ref{eq-stationary}) which are expressed in terms of the Jacobian elliptic functions, for which  we use the standard notations~\cite{AbSt}. While doing this we bear in mind that different functions $\Phi(X)$ which are obtained by the shift of the argument (like, for example, in (\ref{solucion1}) and (\ref{solucion4}) below),  and thus representing the same solution of the homogeneous NLS equation, now will give origin to essentially different solutions of Eq. (\ref{eq-physical}), because the latter are composed of the two periodic  periodic factors steaming from the structure of the periodic potential (they are described by the function $\rho(x)$) and from the NLS solutions mentioned above, (see Eq. (\ref{psi-from-Phi})).

We start with the simplest solutions of Eq. (\ref{eq-physical}) for $g(x)<0$ (i.e. $G=-1$). They are
\begin{subequations}
\begin{eqnarray}
\label{solucion1}
\phi^{(1)}(x)&=&\sqrt{2}k\nu (1+\alpha\cos(2x))\cn(\nu X,k), \quad \nu^{2}=\frac{E}{1-2k^{2}},
 \\
\label{solucion2}
\phi^{(2)}(x)&=&\sqrt{2}\nu (1+\alpha\cos(2x))\dn(\nu X,k),  
\quad \nu^2=\frac{E}{k^2-2},
\\
\label{solucion3}
\phi^{(3)}(x)&=&\sqrt{2(1-k^{2})}\nu (1+\alpha\cos(2x))\frac{1}{\dn(\nu X,k)}, \quad \nu^{2}=\frac{E}{k^{2}-2},
\\
\label{solucion4}
\phi^{(4)}(x)&=&\sqrt{2(1-k^{2})}k\nu(1+\alpha\cos(2x))\frac{\sn(\nu X,k)}{\dn(\nu X,k)},  
\quad \nu^2=\frac{E}{1-2k^2}, 
\end{eqnarray}
\end{subequations}
These are real functions, what determines the regions of the parameters where they are valid. In particular, taking into account that the elliptic modulus $k\in[0,1]$ we have that $\phi^{(2)}(x)$ and $\phi^{(3)}(x)$ are valid only for $E<0$, i.e. these are solutions belonging to the semi-infinite gap, while $\phi^{(1)}(x)$ and $\phi^{(4)}(x)$ belong to the gap $E<0$ and $k>1/\sqrt{2}$ (see Fig.~\ref{Math}(a)) or to the band $E>0$ and $k<1/\sqrt{2}$ (see Figs.~\ref{Math}(c) and (d)). All the solutions bifurcate from the linear Bloch state (\ref{Bloch_10}) recovered at $E=0$.
 
The simplest solutions for $g(x)>0$ ($G=1$) read
\begin{subequations}
\begin{eqnarray}
\label{solucion5}
\phi^{(5)}(x)&=&\sqrt{2}k\nu(1+\alpha\cos(2x))\sn(\nu X,k), \quad \nu^{2}=\frac{E}{1+k^{2}},
\\
\label{solucion6}
\phi^{(6)}(x)&=&\sqrt{2}k\nu (1+\alpha\cos(2x))\frac{\cn(\nu X,k)}{\dn(\nu X,k)}, \quad \nu^{2}=\frac{E}{1+k^{2}}.
\end{eqnarray}
\end{subequations}
As it is clear these solutions are valid for $E>0$. 
  
In the above formulas (\ref{solucion1}) -- (\ref{solucion6}), $X(x)$ is defined in (\ref{eq-sol}), and $\rho(x)$ is given by (\ref{eleccion1}).

Let us now consider the pair of solitons (\ref{solucion5}), (\ref{solucion6}) (or alternatively, the pair (\ref{solucion1}), (\ref{solucion4})) in the "linear" limit $k\to 0$. They readily give two eigenstates of the potential $v(x)$
\begin{subequations}
\label{Bloch_exact}
\begin{eqnarray}
\varphi_1&=&(1+\alpha\cos(2x))\sin\left(\sqrt{E}X\right),\\
 \varphi_2&=&(1+\alpha\cos(2x))\cos\left(\sqrt{E}X\right). 
\end{eqnarray}
\end{subequations}
These are linearly independent solutions. As it is known (see~\cite{MW} for the details) the coexistence of such solutions occurs only if one of the gaps is closed.  Thus, in $\alpha$-dependence of the spectrum of the potential $v(x)$ given by (\ref{v_per}), (\ref{cp}) at $E>0$ there must exits points at which the first lowest gap is closes and through which the chemical potential $\mu$ given by  (\ref{cp}) passes. This is exactly what we observed in panels (c) and (d) of Fig.~\ref{Math}. The existence of such points at sufficiently large $\alpha$ steams from the fact that for $E>0$ the chemical potential infinitely grows with $\alpha$ while the minima of the periodic potential tend to $-\infty$.  

Other exact eigenstates of the potential $v(x)$ given by (\ref{v_per}) can be obtained by considering the linear limit of the solutions
  solutions (\ref{solucion3}) and (\ref{solucion4}) which corresponds to $k\to 1$. In this way we obtain a pair of two "unstable" solutions
  \begin{subequations}
\begin{eqnarray}
	\varphi_3&=&(1+\alpha \cos(2x))\cosh\left(\sqrt{-E}X\right),
	\\ 
	\varphi_4&=&(1+\alpha \cos(2x))\sinh\left(\sqrt{-E}X\right).
\end{eqnarray}
\end{subequations}
 As it is clear the combinations $\varphi_3\pm\varphi_4$ supplied by the proper constant factor provide the asymptotics of the solitary wave $\phi_s(x)$ given by (\ref{eq-sol}) at $x\to\mp\infty$.
  

%
%

Taking into account that $g(x)$ is sign definite, simple stability analysis of the presented solutions can be performed following \cite{VarPer2}: since $\phi^{(2)}(x)>0$ and $\phi^{(3)}(x)>0$ one verifies that they are linearly unstable, while the stability of the solutions $\phi^{(1)}(x)$, $\phi^{(4)}(x)$, $\phi^{(5)}(x)$ and $\phi^{(6)}(x)$ is left undetermined. 
 
Finally, we consider the limit $k\to1/\sqrt{2}$ of the solutions (\ref{solucion1}) and (\ref{solucion4}), which corresponds to the limit of large nonlinearity. For this case, the potential $v(x)$ can be viewed as a small and smooth perturbation of the NLS equation, whose nonlinearity is also a slow function of the spatial variable. Thus the mentioned solutions can be viewed as the periodic NLS solutions modulated by the ``envelope" $\varphi_0(x)$.

\section{Physical applications and concluding remarks.}

Let us now turn to the physical applications of the obtained results. This problem arises naturally since the periodic structures we have used, Eqs. (\ref{v_per}) and (\ref{g_per}), being even expressed in elementary functions, are still not feasible for the most experimental settings, where only one or a few laser beams are used (if one bears in mind optical lattices for BEC applications). One thus can pose the question as follows: do the obtained solutions represent satisfactory approximations to some realistic localized modes (like for example the ones found numerically in~\cite{BludKon}) in the models where the periodic coefficients are represented by a few first Fourier harmonics of the potential $v(x,E)$? The present section aims to answer this question.
 
To this end we Fourier expand the functions $v(x)$ and $g(x)$, and introduce the truncated potentials $v_N(x)$ and $g_N(x)$ generated by superpositions of $N$ harmonics:
\begin{eqnarray}
\label{aproximacion1}
v_N(x)=\sum_{n=1}^N v_{n}(\alpha,E)\cos(2nx),
\ 
g_N(x)=\sum_{n=0}^N g_{n}(\alpha)\cos(2nx),  
\end{eqnarray}
where, for $N=2$ 
\begin{subequations}
\begin{eqnarray}
  \label{aproximacion2vb}
v_{1}
(\alpha,E)
&=&-8\frac{1-\sqrt{1-\alpha^{2}}}{\alpha\sqrt{1-\alpha^{2}}}+ E\frac{\alpha(4+\alpha^{2})}{(1-\alpha^{2})^{7/2}},
\\  \label{aproximacion2vc}
 v_{2}
(\alpha,E)
&=&8\frac{(1-\sqrt{1-\alpha^{2}})^{2}}{\alpha^{2}}- E\frac{5\alpha^{2}}{(1-\alpha^{2})^{7/2}},
\end{eqnarray}
\end{subequations}
and
\begin{subequations}
\begin{eqnarray}
\label{aproximacion2ga}
g_{0}(\alpha)&=&-\frac{1}{8}\frac{40\alpha^{2}+8+15\alpha^{4}}{(1-\alpha^{2})^{11/2}},\\  \label{aproximacion2gb}
g_{1}(\alpha)&=&\frac{3\alpha}{4}\frac{8+12\alpha^{2}+\alpha^{4}}{(1-\alpha^{2})^{11/2}},\\ \label{aproximacion2gc}
g_{2}(\alpha)&=&-\frac{21\alpha^{2}}{4}\frac{2+\alpha^{2}}{(1-\alpha^{2})^{11/2}}.
\end{eqnarray}
\end{subequations}

\begin{figure}
\epsfig{file=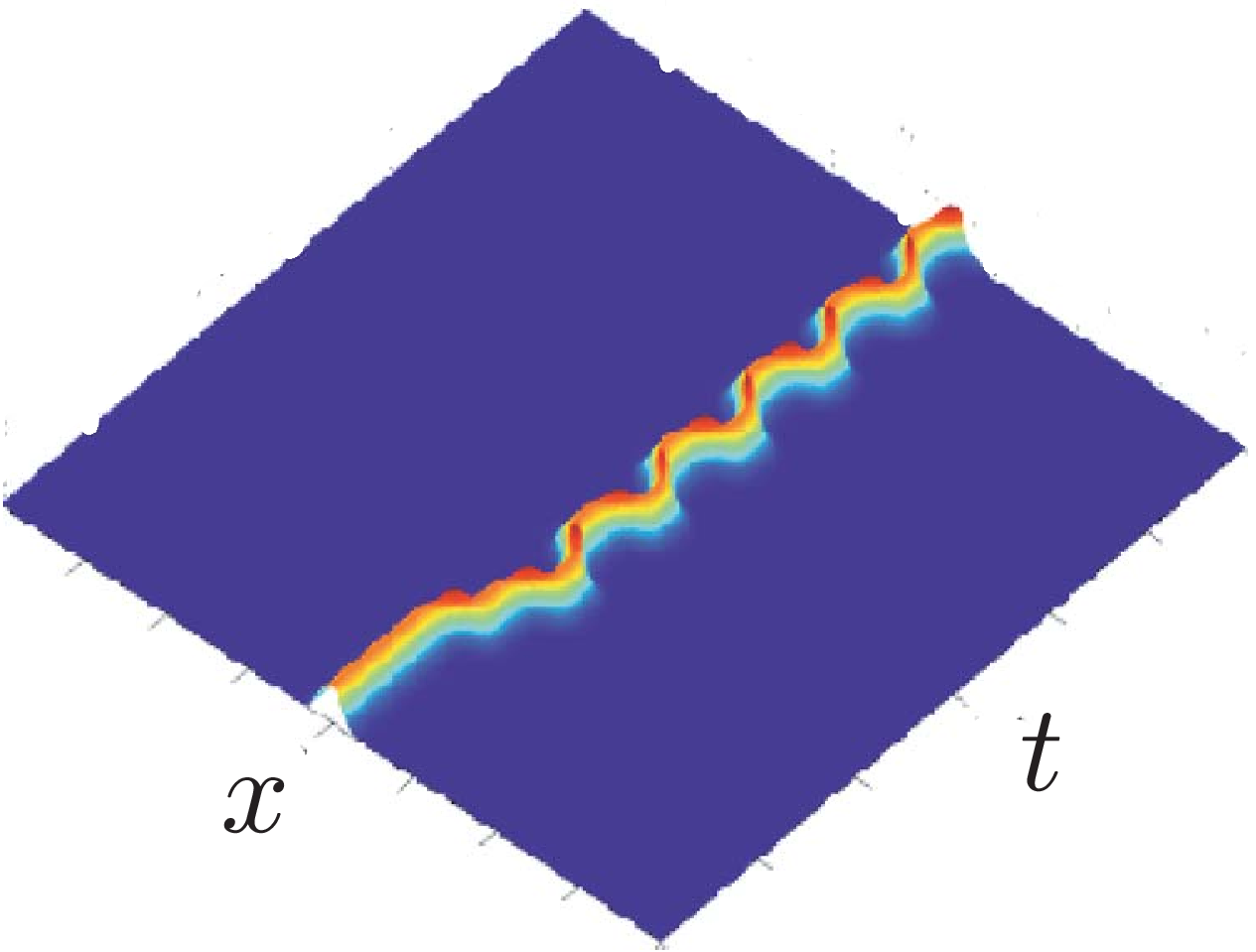,width=8cm}
\epsfig{file=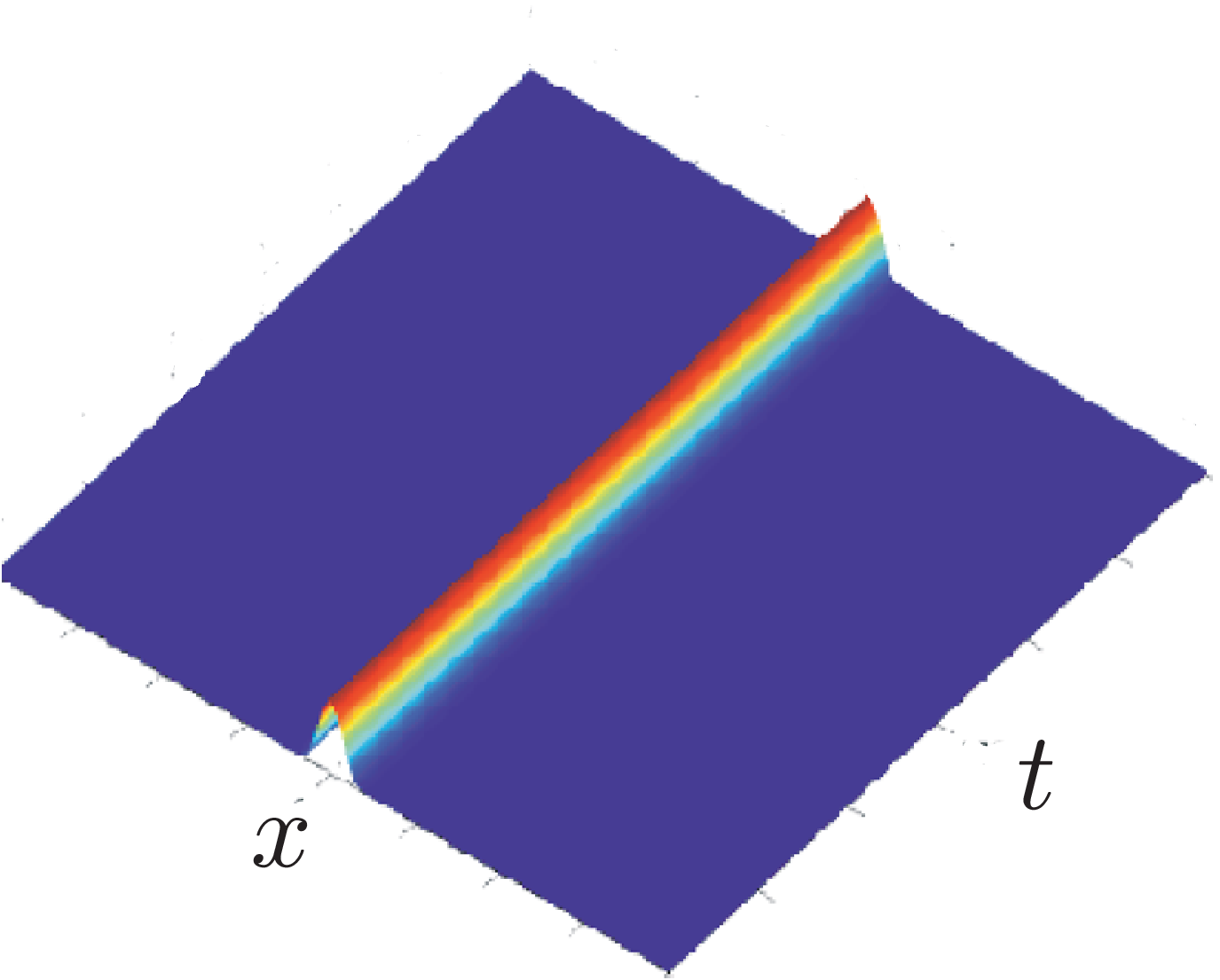,width=8cm}
\caption{Evolution of the solutions (\ref{eq-sol}), with $x\in[-40,40]$, $t\in [0,1500]$, for $E=-0.5$ and $\alpha=-0.1$ (a) using one harmonic (b) two harmonics. For the case (a), the solution is oscillatory. For the case (b) the solution is stable. \label{figurasfourier1}} 
\end{figure}

Next we use the Eqs. (\ref{aproximacion1}), (\ref{aproximacion2vb}), (\ref{aproximacion2vc}) and (\ref{aproximacion2ga}), (\ref{aproximacion2gb}), (\ref{aproximacion2gc}) to approximate the functions $v(x)$ and $g(x)$, given by Eqs. (\ref{g_per}) and (\ref{v_per}), respectively. Thus, for example, for $\alpha=-0.1$ and $E=-0.5$, which are values where the solutions of Eq. (\ref{eq-physical}) are stable, (see Fig. \ref{autovalores}(a)), we obtain
\begin{subequations}
\begin{eqnarray}
g&=&-1.1099-0.6436\cos(2x)-0.1115\cos(4x),\\
v&=&0.6107\cos(2x)+0.0561\cos(4x).
\end{eqnarray}
\end{subequations}
Finally, we have simulated numerically the dynamics of the wave packet with the initial profile given by the solution (\ref{eq-sol}), and described by the evolution equation (\ref{eq-physical1}) with the potentials $v_N$ and $g_N$ with $N=1,2$ instead of $v$ and $g$, respectively. Some typical  results are shown in Fig. \ref{figurasfourier1}. One observes that while the explicit analytical solution  (\ref{eq-sol}) is not a satisfactory approximation for the harmonic potentials: in Fig.~\ref{figurasfourier1}(a) one observes oscillatory behavior of the mode. Using a two-harmonic approximation for the potential results in a very stable behavior of the solution.

\begin{figure}
\epsfig{file=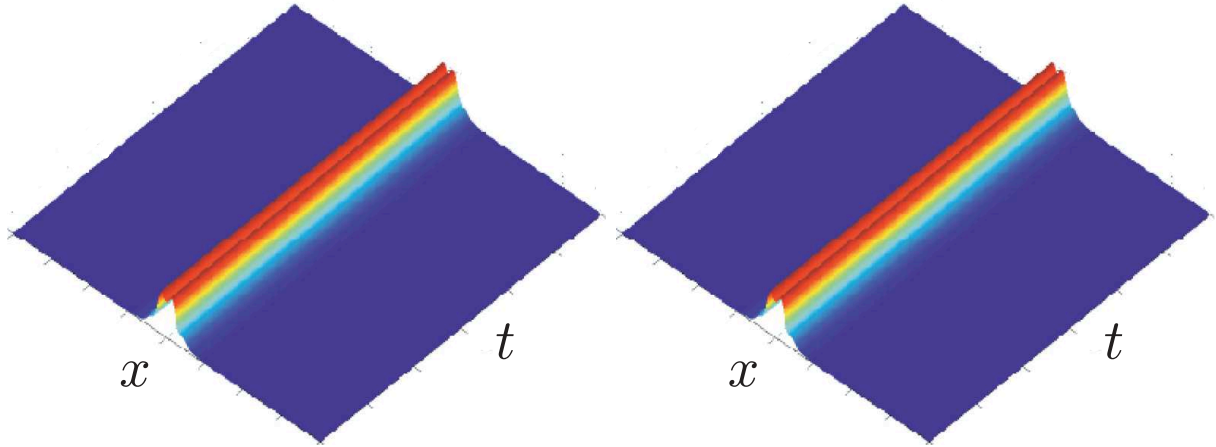,width=8cm}
\epsfig{file=solucion1E_01a_01.eps,width=8cm}
\caption{Evolution of the solutions (\ref{eq-sol}), with $x\in[-40,40]$, $t\in [0,1500]$, for $E=-0.1$ and $\alpha=-0.1$ where the linear and nonlinear lattices are approximated by   using only one harmonic (a) and  two harmonics (b). For both cases, the solution is stable.  
 \label{figurasfourier2}}
\end{figure}

It turns out, that for some specific domains of the parameters the obtained solutions represent fairly good approximations even for the harmonic potentials. An example of such a situation is shown in Fig.~\ref{figurasfourier2} where we have choses  $E=-0.1$ and $\alpha=-0.1$: no visible difference exist in the dynamical regime for the monochromatic lattices (Fig.~\ref{figurasfourier2}(a)) and the lattices in a form of superposition of the two harmonics (Fig.~\ref{figurasfourier2}(b)).

Both Figs. \ref{figurasfourier1} and \ref{figurasfourier2} were obtained by direct numerical simulations of Eq. (\ref{eq-physical1}), using as initial data in the evolution in time the solution (\ref{eq-sol}).



To conclude, in this paper, by using similarity transformations we have constructed explicit localized solutions of the nonlinear Schr\"odinger equation with linear periodic potential and spatially periodic nonlinearities. We have studied such solutions and their linear stability. We also have calculated periodic solutions of the inhomogeneous nonlinear Schr\"odinger equation and have shown that they reveal some properties of the underline linear lattices, providing exact analytical expressions for the respective Bloch waves. 
 
\section*{Acknowledgements}
We would like to thank J. Cuevas for discussions.
VVK acknowledges support from Ministerio de Educaci\'on y Ciencia (MEC, Spain) under grant SAB2005-0195 and PCI08-0093. 
This work has been partially supported by grants  FIS2006-04190 (MEC), POCI/FIS/56237/2004 (FCT -Portugal- and European program FEDER) and PCI-08-0093 (Junta de Comunidades de Castilla-La Mancha, Spain).

\appendix

\section{Examples of models allowing exact solutions}

In this appendix, we present two more general models that we used in Eqs. (\ref{v_per})  and (\ref{g_per}).
The first model corresponds to the choice
\begin{equation}
\rho(x)=(1+\alpha\cos(2 x))^{p},
\end{equation} 
with $p\in\mathbb{R}$.
So, it is clear that 
\begin{equation}
g(x)=\frac{G}{(1+\alpha\cos(2 x))^{6p}},
\end{equation}
and  the external potential becomes
\begin{multline}
v(x)=-4\alpha p\left(1+\alpha\cos(2 x)\right)^{-2}\left[\alpha(1-p)+\cos(2 x)+p\alpha\cos^{2}(2 x)\right]\\ 
- \frac{E}{(1+\alpha\cos(2 x))^{4p}}+\mu.
\end{multline}
It is clear that when $p=1$, we recover the expressions (\ref{v_per}) and (\ref{g_per}).

The second model corresponds to 
\begin{equation}
g(x)= \eta\left(\dn(\xi,k)\right)^{p},
\end{equation}
with $\eta$ a constant and $\dn$ the Jacobi elliptic function. We can calculate the external potential $v(x)$ and obtain
\begin{multline}
v(x)=\frac{p}{6}\left[\left(1-\tfrac{p}{6}\right)(\dn(\xi,k))^2\right]+\frac{p}{6}\left[\left(k^2-1\right)\left(1+\frac{p}{6}\right)\frac{1}{(\dn(\xi,k))^{2}}+p\left(\frac{1}{3}-\frac{k^2}{6}\right)\right]\\ 
-E\left(\frac{\eta}{G})^{2/3}(\dn(\nu\xi,k)\right)^{2p/3}+\mu.
\end{multline}
One can obtain the nonlinearity $g(x)$ and the potential $v(x)$, according to the value of $p$.


\begin{thebibliography}{10}



\bibitem{BK_rev} V.A. Brazhnyi and V.V. Konotop, Theory of nonlinear matter waves in optical lattices, Mod. Phys. Lett. B, {\bf 18}, 627 (2004).  

\bibitem{reviews} 
O. Morsch, and M. Oberthaler, Dynamics of Bose-Einstein condensates in optical lattices, Rev. Mod. Phys. {\bf 78}, 179 (2006).

\bibitem{BKK} V. A. Brazhnyi, V. V. Konotop, and V. Kuzmiak, Dynamics of matter solitons in weakly modulated optical lattices, Phys. Rev. A, {\bf 70}, 043604 (2004).

\bibitem{BMS} B. B. Baizakov, B. A. Malomed, and M. Salerno, Multidimensional solitons in a low-dimensional periodic potential, Phys. Rev. A, {\bf 70}, 053613 (2004).

\bibitem{BKPG} V. A. Brazhnyi, V. V. Konotop, and V. M. P\'erez-Garc\'ia, Driving defect modes  of Bose-Einstein condensates in optical lattices, Phys.Rev. Lett., {\bf 96}, 060403 (2006).

\bibitem{BKPG1} V. A. Brazhnyi, V. V. Konotop, and V. M. P\'erez-Garc\'ia, Defect modes of a Bose-Einstein condensate in an optical lattice with a localized impurity, Phys. Rev. A, {\bf 74}, 023614 (2006).

\bibitem{BKPG2} V. Ahufinger, A. Mebrahtu, R. Corbal\'an, and A. Sanpera, Quantum switches and quantum memories for matter-wave lattice solitons, New J. Phys., {\bf 9}, 4 (2007). 
 
%
%
%

%
%
%
%
\bibitem{SM} H. Sakaguchi and B. A. Malomed, Matter-wave solitons in nonlinear optical lattices, Phys. Rev. E, {\bf 72}, 046610 (2005).

\bibitem{Fibich} G. Fibich, Y. Sivan, and M. I. Weinstein, Bound states of NLS equations with a periodic nonlinear microstructure, Physica D, {\bf 217}, 31-57 (2006).

\bibitem{Sivan2}  Y. Sivan, G. Fibich, and M. I. Weinstein, Waves in nonlinear lattices - ultrashort optical pulses and Bose-Einstein condensates, Phys. Rev. Lett., {\bf 97}, 193902 (2006).

 
 
\bibitem{BludKon} Y. Bludov, and V. V. Konotop, Localized modes in arrays of boson-fermion mixtures, Phys. Rev. A,  \textbf{74}, 043616 (2006).

\bibitem{interplay}  Z. Rapti, P.G. Kevrekidis, V.V. Konotop, C.K.R.T. Jones,  Solitary waves under the competition of linear and nonlinear periodic potentials, arXiv:0707.1162.

\bibitem{VarPer} J.C. Bronski, L.D. Carr, B. Deconinck, J.N. Kutz, and K. Promislow, Stability of repulsive Bose-Einstein condensates in a periodic potential, Phys. Rev. E, {\bf 63}, 036612 (2001).

\bibitem{VarPer2}J.C. Bronski, L.D. Carr, R. Carretero-Gonzalez, B. Deconinck, J.N. Kutz, and K. Promislow, Stability of attractive Bose-Einstein condensates in a periodic potential, Phys. Rev. E, {\bf 64}, 056615 (2001). 

\bibitem{JVV}{{J. Belmonte-Beitia, V. M. P\'erez-Garc\'ia and V. Vekslerchik}, {Modulational instability, solitons and periodic waves in models of quantum degenerate Boson-Fermion mixtures}, Chaos, Solitons and Fractals, {\bf 32}, (2007) 1268-1277. }

\bibitem{CSF1}{{ B. Li and Y. Chen}, { On exact solutions of the nonlinear Schr\"odinger equations in optical fiber}, Chaos, Solitons and Fractals, {\bf 21}, (2004) 241-247.}

\bibitem{CSF2}{{ J-M. Zhu and Z-Y. Ma}, { Exact solutions for the cubic-quintic nonlinear Schr\"odinger equation}, Chaos, Solitons and Fractals, {\bf 33}, (2007) 958-964.}

\bibitem{CSF3}{{ S-D Zhu}, { Exact solutions for the high-order dispersive cubic-quintic nonlinear Schr\"odinger equation by the extended hyperbolic auxiliary equation method}, Chaos, Solitons and Fractals, {\bf 34} (2007) 1608-1612.}


\bibitem{Serkin} H.-H. Chen and C.-S. Liu, Phys. Rev. Lett., Solitons in nonuniform media, {\bf 37}, 693 (1976).

\bibitem{Serkin1} V. V. Konotop, O. A. Chubykalo, and L. V\'azquez, Dynamics and interactions of solitons on an integrable inhomogeneous lattice, Phys. Rev. E, {\bf 48}, 564 (1993).

\bibitem{Serkin2} V. N. Serkin and A. Hasegawa, Novel soliton solutions of the nonlinear Schr\"odinger equation model, Phys. Rev. Lett., \textbf{85}, 4502 (2000).

\bibitem{Serkin3} Z. X. Liang, Z. D. Zhang, W. M. Liu,  Dynamics of a Bright Soliton in Bose-Einstein Condensates with Time-Dependent Atomic Scattering Length in an Expulsive Parabolic Potential,  Phys. Rev. Lett., {\bf 93}, 152301 (2004).

\bibitem{Serkin4} S. Chen and L. Yi, Chirped self-similar solutions of a generalized nonlinear Schr\"odinger equation model, Phys. Rev. E, {\bf 71}, 016606 (2005).

\bibitem{Serkin5} V. M. P\'erez-Garc\'{\i}a, P. J. Torres, and V. V. Konotop, Similarity transformations of nonlinear Schr\"odinger equations with time varying coefficients: Exact results, Physica D, {\bf 221}, 31 (2006).

\bibitem{Serkin6}  V. N. Serkin, A. Hasegawa, and T. L. Belyaeva, Nonautonomous solitons in external potentials,  Phys. Rev. Lett., {\bf 98}, 074102 (2007).  
 
\bibitem{exact} J. Belmonte-Beitia, V. M. P\'erez-Garcia, V. Vekslerchik, and P. J. Torres, Lie symmetries and solitons in nonlinear systems with spatially inhomogeneous nonlinearities, Phys. Rev. Lett., {\bf 98}, 064102 (2007). 

\bibitem{exact2} J. Belmonte-Beitia, V. M. P\'erez-Garcia, V. Vekslerchik, and P. J. Torres, Lie symmetries, qualitative analysis and exact solutions of nonlinear Schr\"odinger equations with inhomogeneous nonlinearities, Discrete and Continuous Dynamical Systems B, {\bf 9}, 2 (2008).

\bibitem{Sivan} Y. Sivan, G. Fibich, N. K. Efremidis, S. Bar-Ad, Analytic theory of narrow lattice solitons, arXiv:0707.1589.


\bibitem{AKS} G. L. Alfimov, V. V. Konotop, and M. Salerno, Matter solitons in Bose-Einstein Condensates with optical lattices, Europhys. Lett., {\bf 58}, 7 (2002).

\bibitem{AbSt}  D.K. Lawden, {\em Elliptic Functions and applications}, Springer-Verlag, New York Inc., (1989).










\bibitem{MW} W. Magnus and S. Winkler, {\em Hill's Equation}, Dover Publications, Inc. New York, (1979).

\end{thebibliography}
\end{document}